\documentclass[twoside,twocolumn,9pt]{article}
\usepackage{extsizes}
\usepackage[super,sort&compress,comma]{natbib}
\usepackage[version=3]{mhchem}
\usepackage[left=1.5cm, right=1.5cm, top=1.785cm, bottom=2.0cm]{geometry}
\usepackage{balance}
\usepackage{widetext}
\usepackage{times}
\usepackage{newtxtext,newtxmath}
\usepackage{eqnarray,amsmath}
\usepackage{sectsty}
\usepackage{graphicx}
\usepackage{booktabs}
\usepackage{lastpage}
\usepackage[format=plain,justification=raggedright,singlelinecheck=false,font={stretch=1.125,small,sf},labelfont=bf,labelsep=space]{caption}
\usepackage{float}
\usepackage{fancyhdr}
\usepackage{fnpos}
\usepackage[english]{babel}
\usepackage{array}
\usepackage{droidsans}
\usepackage{charter}
\usepackage[T1]{fontenc}
\usepackage[usenames,dvipsnames]{xcolor}
\usepackage{setspace}
\usepackage[compact]{titlesec}
\usepackage{bm}
\usepackage{multicol}

\usepackage{epstopdf}

\definecolor{cream}{RGB}{222,217,201}

\begin{document}



\makeFNbottom
\makeatletter

\def\red{\textcolor{red}}

\def\Gr{{\rm \Gamma}}
\newcommand{\be}{\begin{equation}}
\newcommand{\ee}{\end{equation}}
\newcommand{\bea}{\begin{eqnarray}}
\newcommand{\eea}{\end{eqnarray}}

\renewcommand\LARGE{\@setfontsize\LARGE{15pt}{17}}
\renewcommand\Large{\@setfontsize\Large{12pt}{14}}
\renewcommand\large{\@setfontsize\large{10pt}{12}}
\renewcommand\footnotesize{\@setfontsize\footnotesize{7pt}{10}}
\makeatother

\renewcommand{\thefootnote}{\fnsymbol{footnote}}
\renewcommand\footnoterule{\vspace*{1pt}%
\color{cream}\hrule width 3.5in height 0.4pt \color{black}\vspace*{5pt}}
\setcounter{secnumdepth}{5}

\makeatletter
\renewcommand\@biblabel[1]{#1}
\renewcommand\@makefntext[1]%
{\noindent\makebox[0pt][r]{\@thefnmark\,}#1}
\makeatother
\renewcommand{\figurename}{\small{Fig.}~}
\sectionfont{\sffamily\Large}
\subsectionfont{\normalsize}
\subsubsectionfont{\bf}
\setstretch{1.125} 
\setlength{\skip\footins}{0.8cm}
\setlength{\footnotesep}{0.25cm}
\setlength{\jot}{10pt}
\titlespacing*{\section}{0pt}{4pt}{4pt}
\titlespacing*{\subsection}{0pt}{15pt}{1pt}

\fancyfoot{}
\fancyfoot[LE]{\footnotesize{\sffamily{\hspace{2pt}\thepage}}}
\fancyfoot[RO]{\footnotesize{\sffamily{\hspace{2pt}\thepage}}}

\fancyhead{}
\renewcommand{\headrulewidth}{0pt}
\renewcommand{\footrulewidth}{0pt}
\setlength{\arrayrulewidth}{1pt}
\setlength{\columnsep}{6.5mm}
\setlength\bibsep{1pt}

\makeatletter
\newlength{\figrulesep}
\setlength{\figrulesep}{0.5\textfloatsep}

\newcommand{\topfigrule}{\vspace*{-1pt}%
\noindent{\color{cream}\rule[-\figrulesep]{\columnwidth}{1.5pt}} }

\newcommand{\botfigrule}{\vspace*{-2pt}%
\noindent{\color{cream}\rule[\figrulesep]{\columnwidth}{1.5pt}} }

\newcommand{\dblfigrule}{\vspace*{-1pt}%
\noindent{\color{cream}\rule[-\figrulesep]{\textwidth}{1.5pt}} }

\makeatother

\twocolumn[
  \begin{@twocolumnfalse}
\sffamily
\begin{tabular}{m{0.0cm} p{16.0cm} }

\quad &
\noindent\LARGE{\textbf{Behaviour of the model antibody fluid constrained by rigid spherical obstacles. Effects of the obstacle--antibody attraction}} \\
\vspace{0.3cm} & \vspace{0.3cm} \\
& \noindent\large{Taras Hvozd$^1$, Yurij V. Kalyuzhnyi$^1$, Myroslav Holovko$^1$, Vojko Vlachy$^2$} \\
\quad & \textit{$^1$~Institute for Condensed Matter Physics of the National Academy of Sciences of Ukraine 1 Svientsitskii St., Lviv, Ukraine 79011;} \\ 
&\textit{ E-mail: tarashvozd@icmp.lviv.ua,$\;\;$yukal@icmp.lviv.ua, $\;\; $holovko@icmp.lviv.ua} \\ 
\quad & \textit{$^2$~Faculty of Chemistry and Chemical Technology, University of Ljubljana, Ve\v{c}na pot 113, SI--1000 Ljubljana, Slovenia. } \\ 
&\textit{E-mail: vojko.vlachy@fkkt.uni-lj.si} \\  
\vspace{0.3cm} & \vspace{0.3cm} \\
\quad 
& \noindent\normalsize
{This study is concerned with behaviour of fluid of monoclonal antibodies (mAbs) when trapped in a confinement represented by rigid spherical obstacles that attract proteins. The antibody molecule is depicted as an assembly of seven hard spheres, organized to resemble Y shaped molecule. The antibody has two Fab and one Fc domains located in the corners of letter Y. In this calculation, only the Fab-Fab and Fab-Fc attractive pair interactions are effective. The confinement is formed by the randomly distributed hard-spheres fixed in space. The spherical obstacles, besides the size exclusion, also interact by the Yukawa attractive interaction with with each bead of the antibody molecule. We applied the combination of the scaled-particle theory, Wertheim's thermodynamic perturbation theory and the Flory-Stockmayer theory to calculate: (i) the liquid-liquid phase separation, and (ii) the percolation threshold. All these quantities were calculated as functions of the strength of the attraction between the monoclonal antibodies, and monoclonal antibodies and obstacles. The conclusion is that while the hard-sphere obstacles decrease the critical density as also, the critical temperature of the mAbs fluid, the effect of the protein-obstacle attraction is more complex. Adding the attraction to obstacle-mAbs interaction first increases the wideness of the temperature-density envelope. However, with the further increase of the obstacle-mAbs attraction intensity we observe reversal of the effect, the temperature-density curves become narrower. At some point, depending on the AC=BC interaction, the situation is observed where two different temperatures have the same fluid density (reentry point). In all the cases shown here the critical point decreases below the value for the neat fluid, but the behaviour with respect to increase of the strength of obstacle-mAbs attraction is not monotonic.}
\end{tabular}

 \end{@twocolumnfalse} \vspace{0.6cm}

]

\renewcommand*\rmdefault{bch}\normalfont\upshape
\rmfamily
\section*{}
\vspace{-1cm}


\section{Introduction}

Protein properties are most often studied in diluted 
aqueous salt solutions, what is in sharp contrast with the cellular environment, which is densely packed with macromolecules such as nucleic acids, other protein species, 
etc. In other words actual (in ``vivo'') conditions are very different from those in which our experimental and theoretical studies are most often performed~\cite{Zimm1993,hall2003,Elcock2010}.  
A recent tutorial paper dealing with the effects of crowding on biochemical systems is due to Silverstein and Slade ~\cite{Silver2019} and it contains some important newer references. 

It is generally believed that the effect of crowding can be divided into two major components, one is the excluded volume (``hard'') component, which was examined in many papers and the ``soft'' attractive interactions, which are considerably less studied. The two components oppose each other so the resulting effect may not be strong. In the previous paper~\cite{Hvozd2020} we, following other authors, \cite{Zhou2008,Elcock2010,Qin2017,Cho2012} examined effects of the hard--sphere obstacles on properties of the model monoclonal antibody (mAb) fluid. The physico-chemical stability of these special proteins (mAbs) used as bio-therapeutics,  has been reviewed recently~\cite{Lebasle2020}.

It has recently emerged that membrane--less organelles (``protein droplets'') formed in the cell via reversible liquid--liquid phase separation are associated with important cellular functions~\cite{Shin2017,Hyman2012,Banani2017,Nguemaha2018a,Alberti2019,Mcswiggen2019}. These ``droplets'' 
are reversible and may disappear upon adding the electrolyte, changing the $p$H or temperature. The liquid--liquid phase transition is the main property examined in this study. In particular we are interested how the ``bystander'' species (polymers, proteins, etc. are in literature also called the regulatory components) present in the system influence the shape of the liquid--liquid phase separation curve. It was confirmed for the excluded-volume effects of the obstacles to favor an emergence of the protein--rich phase, while attractive interactions shall partly counteract the effect of hard-core repulsion, decreasing the colloidal stability of proteins~\cite{Sarkar2014}.
In this context we need to mention the paper of Qin and Zhou~\cite{Qin2017}, reviewing studies of the protein liquid-liquid phase separation under cell-like conditions. One important conclusion of their work is that the phase boundaries may be shifted upon addition of the crowding agent. 

In one of the previous works we analyzed~\cite{Kastelic2016b} (a similar study has been performed by Dorsaz et al ~\cite{Dorsaz2011SM}) experimental data for the liquid--liquid  phase separation~\cite{Benedek2010} in $\gamma$D--$\beta$B1 crystallin mixtures. The effect of the second protein (regulatory component) added to the solution of the first one was to decrease the critical temperature of the mixture and to make the phase boundary, i.e. the T--$\rho$ envelope, narrower. In a more recent study we proposed the calculation where model monoclonal antibodies were distributed within an array of obstacles (also called matrix) formed by immobile (fixed) hard spheres. 
We can assume that obstacles represent the molecules, having much lower mobility than the protein component. In literature such examples are known as partly quenched systems (some degrees of freedom are quenched and the others are free), being theoretically examined by Madden, Glandt and many others~\cite{Madden-Glandt1988,Madden1992}. For review of such studies see, for example Ref.~\cite{Luksic2011}. While in the previous study the obstacle particles were considered to be inert hard spheres, in the present calculation the obstacles have an ability to attract model proteins. For this purpose the obstacles are decorated by attractive sites to which the molecule of mAbs can bind. We will show that a presence of attractive protein-obstacle interactions substantially modifies the liquid-liquid phase diagram, making it much richer.

\section{Modeling antibody molecules in Yukawa hard-sphere matrix}

The 7-bead model of the antibody molecule has already been studied in previous papers \cite{Kalyuzhnyi2018,Kastelic2017,Kastelic2018,Hvozd2020}. The molecules of immunoglobulin are represented by seven tangentially bonded equal size hard-sphere beads, forming completely flexible three-arm star-like structure, assuming approximately the Y-shape on the average. In addition each of the terminal beads have one square-well off-center site, located on the surface. Two upper forks represent
fragment antigen-binding regions (Fab and Fab'), here denoted as A and B, while the base represents the fragment crystallizable region (Fc), denoted by C. Porous medium is modelled as the matrix of the hard-sphere obstacles randomly placed in the configuration of the hard-sphere fluid, quenched at equilibrium. The 
attractive Yukawa  interaction takes place between the matrix particles and each bead {(center to center)} of the antibody molecule. 
The number densities of the antibody molecules and matrix obstacles are $\rho_1$ and $\rho_0$, respectively. 

The pair potential $U_{KL}(12)$, which is acting between the terminal beads of two different
immunoglobulin molecules bearing the sites of the type $K$ and $L$, is
\be
U_{KL}(12)=U_{hs}(r)+U_{KL}^{(as)}(12),
\label{U}
\ee
where $U_{hs}(r)$ is the hard-sphere potential,
\be
U^{(as)}_{KL}(12)=U^{(as)}_{KL}(z_{12})=\left\{
\begin{array}{rl}
	-\epsilon^{(as)}_{KL}, & {\rm for}\;z_{12}\le\omega\\
	0, & {\rm otherwise}
\end{array}
\right.,
\label{UKL}
\ee
$1$ and $2$ denote position and orientation of the two beads in question, $K,L$ take the 
values $A,B$ and $C$, $z_{12}$ is the distance between the sites $K$ and $L$,
$\epsilon^{(as)}_{KL}$ and $\omega$ are the depth and width of the site-site square-well potential.
For the Yukawa potential $U_{01}^{(Y)}(r)$ we have
\be
U_{01}^{(Y)}(r)=-{\epsilon_{01}^{(Y)}\sigma_{01}\over r}\exp{\left[-\alpha\left(r-\sigma_{01}
	\right)\right]},
\label{UY}
\ee
where $r$ is the distance between the matrix hard sphere and corresponding bead center of the
immunoglobulin molecule. Also, $\sigma_{01}=(\sigma_0+\sigma_1)/2$, where $\sigma_0$ and $\sigma_1$ are 
the hard-sphere diameters of the matrix obstacles and immunoglobulin beads, respectively.

\section{Theory}

Our approach uses a combination of Wertheim's thermodynamic perturbation theory (TPT) for associating fluids  \cite{wertheim1986,wertheim1986a,Jackson1988}, scaled particle theory (SPT) for the porous media~\cite{Holovko2013,Kalyuzhnyi2014a,Hvozd2017,Hvozd2018,Hvozd2020}, and replica Ornstein-Zernike (ROZ) equations~\cite{Madden-Glandt1988,Madden1992}. The latter two theories are used to describe the properties of the reference system. According to the TPT, Helmholtz free energy of the model can be written as a sum of the two terms, the reference system term $A_{ref}$ and the term $\Delta A_{as}$ describing contribution due to associative interaction,
\be
A=A_{ref}+\Delta A_{as}.
\label{Arefass}
\ee
The reference system is represented by the fluid of non-associating antibody molecules 
($\epsilon_{KL}=0$) confined by the matrix of Yukawa hard-sphere obstacles. The corresponding 
expression for Helmholtz free energy $A_{ref}$ is
$$
{\beta A_{ref}\over V}=\rho_1\left[ 7\ln{\left(\Lambda_1^3\rho_1\right)}-1 \right]
$$
\be
-6\rho_1\ln{\left[\sigma_1^3\rho_1g_{11}^{(Y)}(\sigma_1^+)\right]}+
{\beta\Delta A_{hs}\over V}+{\beta\Delta A_Y\over V}.
\label{Aref}
\ee
$\Delta A_{hs}$ is the excess free energy of the hard-sphere fluid confined in the hard-sphere matrix and $\Delta A_{Y}$ represent contribution of the Yukawa interaction to the free  energy of the reference system. The former term is calculated using the SPT 
~\cite{Holovko2013,Kalyuzhnyi2014a,Hvozd2017,Hvozd2018,Hvozd2020}, which provides analytical expressions for Helmholtz free energy, chemical potential, and pressure. These expressions were explicitly presented in our previous study \cite{Hvozd2020} yet,
for the sake of completeness, we include them in the Appendix A. 

The contribution of the Yukawa interaction to thermodynamic properties of the reference system was  calculated using ROZ equations supplemented by the Percus-Yevick and Mean Spherical Approximation (MSA) closure relations. ROZ theory was presented and discussed in a number of the previous studies (see, e.g. \cite{Luksic2011}) and we refer the readers to the original publications for details. In addition, in the Appendix B we present the basic equations specialised for the model used here. 

For $\Delta A_{as}$ we have 
\be
{\beta\Delta A_{as}\over V}=\rho_1\left[\ln{\left(X_AX_BX_C\right)}-
{1\over 2}\left(X_A+X_B+X_C+3\right)\right],
\label{Aass}
\ee
where $X_A$, $X_B$ and $X_C$ are the fractions of the particles with the non-bonded sites 
$A$, $B$, and $C$, respectively. These fractions follow from the solution of the following set of equations
\be  
\rho_1X_K\sum_{L=A}^C\Delta_{KL}X_L+X_K-1=0,
\label{mass}
\ee
where
\be
\Delta_{KL}=4\pi g_{11}(\sigma_1^+)\int_{\sigma_{1}}^{\sigma_1+\omega}
{\bar f}_{KL}(r)r^2dr,
\label{Del}
\ee
${\bar f}_{KL}(r)$ is orientation averaged Mayer function for the site-site square-well 
potential
\be
{\bar f}_{KL}(r)=\left(e^{\beta\epsilon_{KL}}-1\right)\left(\omega+\sigma_1-r\right)^2
\left(2\omega-\sigma_1+r\right)/\left(6\sigma_1^2r\right),
\label{fbar}
\ee
$g_{11}(\sigma_1^+)$ is the contact value of the pair distribution function of the
hard-sphere fluid confined in the Yukawa hard-sphere matrix.

Using the above expression for Helmholtz free energy (\ref{Arefass}), the chemical potential of 
the antibody particles and equation of state (pressure) can be calculated via standard 
thermodynamic relations. These quantities are needed to obtain the liquid-liquid phase 
diagram, which follows from the solution of equations representing equilibrium conditions
\be
\mu(T,\rho_1^{(g)})=\mu(T,\rho_1^{(l)}),
\label{mulg}
\ee
\be
P(T,\rho_1^{(g)})=P(T,\rho_1^{(l)}),
\label{Plg}
\ee
where $\rho_1^{(g)}$ and $\rho_1^{(l)}$ are the densities of the the coexisting ``gas'' 
(low density) and ``liquid'' (high density) phases.

Similar as in our previous study\cite{Hvozd2020}, we calculated the percolation threshold line. These calculations are based on the corresponding extension of the Flory-Stockmayer theory proposed for patchy colloid models with identical bonding sites~\cite{Bianchi2007JPCB,Bianchi2008}, later generalized for models with different bonding sites~\cite{Tavares2010PRE,Tavares2010}.

\section{Results and Discussion}

\begin{figure*}[h]
	\includegraphics[keepaspectratio=true,width=0.45\textwidth]{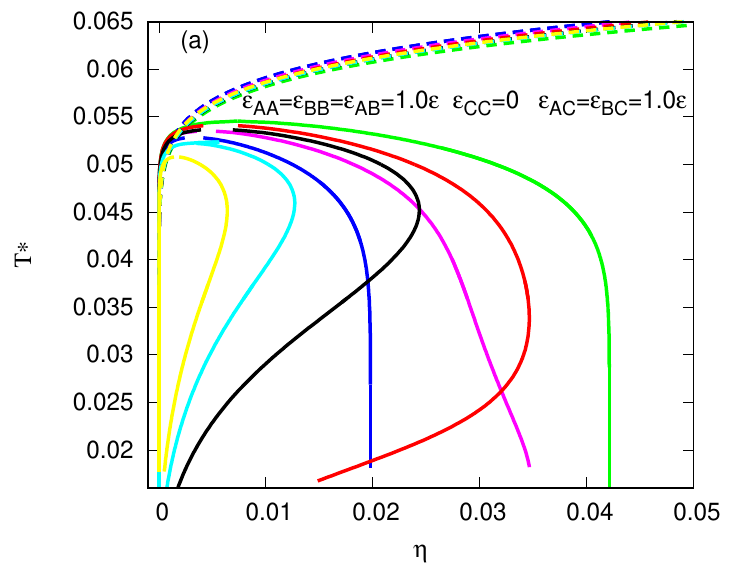} 
	\includegraphics[keepaspectratio=true,width=0.45\textwidth]{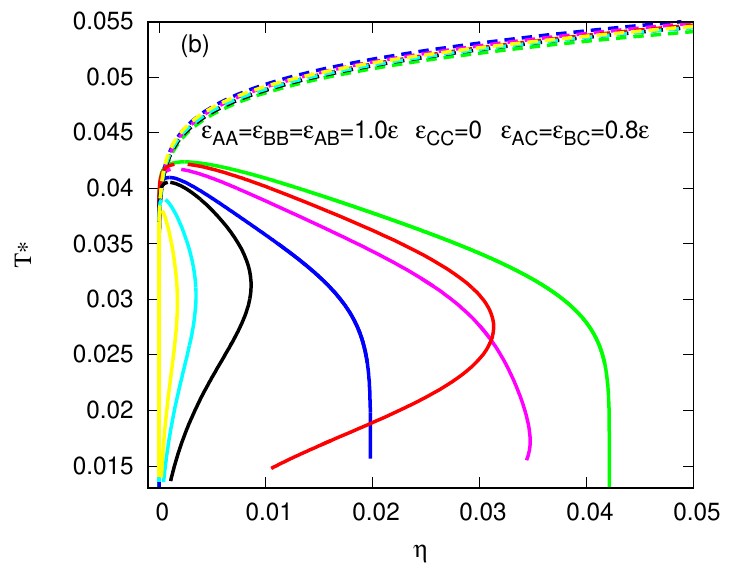}
	
	\includegraphics[keepaspectratio=true,width=0.45\textwidth]{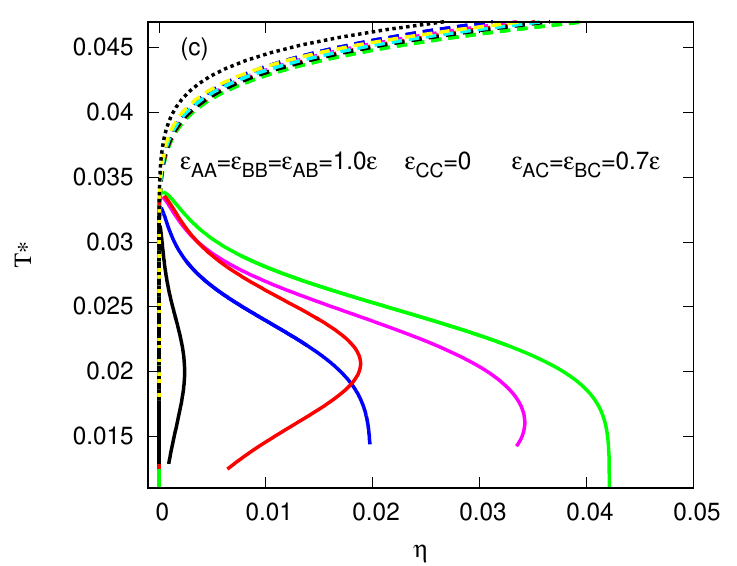}
	\includegraphics[keepaspectratio=true,width=0.45\textwidth]{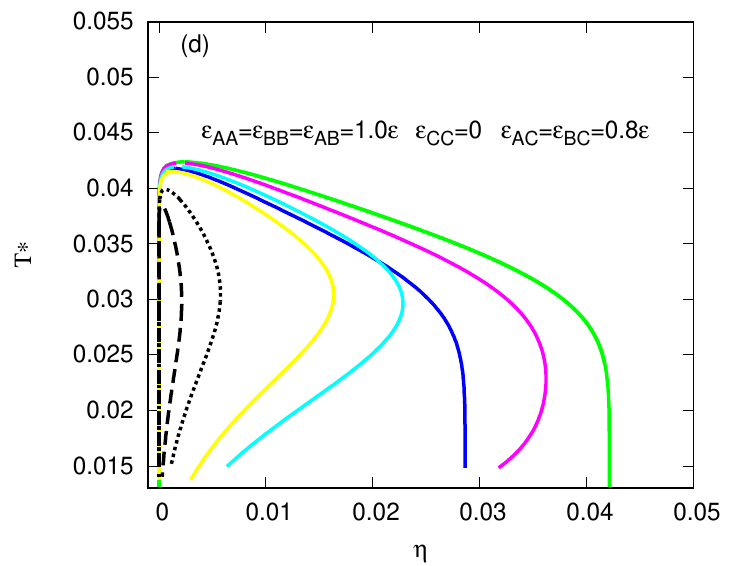}
	\caption{\small Phase diagrams in $T^*$ vs $\eta=\pi\rho_1^*/6$ coordinate frame
		for model of monoclonal antibodies in Yukawa hard-sphere porous media at bonding distance $\omega=0.05\sigma_{1}$ and (a) $\epsilon^{(as)}_{AA}=\epsilon^{(as)}_{BB}=\epsilon^{(as)}_{AB}=\epsilon$, $\epsilon^{(as)}_{CC}=0$, $\epsilon^{(as)}_{AC}=\epsilon^{(as)}_{BC}=1.0\epsilon$, {$\tau=1$}, obstacles packing fraction $\eta_{0}=0.1$ and different strength of the Yukawa interaction: $\epsilon_Y=0$ (blue line), $\epsilon_Y=0.015\epsilon$ (purple line), $\epsilon_Y=0.03\epsilon$ (red line), $\epsilon_Y=0.06$ (black solid line), $\epsilon_Y=0.08\epsilon$ (cyan line), $\epsilon=0.1\epsilon$ (yellow line). In all plots the green lines denote the result for the fluid with no obstacles present ($\eta_0=0$). (b) The same as in Fig.1a but for $\epsilon_{AC}=\epsilon_{BC}=0.8\epsilon$. (c) The same as for Fig. 1a and 1b but for $\epsilon_{AC}=\epsilon_{BC}=0.7\epsilon$. (d) The same as for Fig. 1b but for $\tau=\sigma_{1}/\sigma_0$=0.5. The color code is: $\epsilon_Y=0$ (blue line), $\epsilon_Y=0.03547\epsilon$ (purple line corresponds to purple line in Fig. 1b or $\epsilon_Y=0.015\epsilon$), $\epsilon_Y=0.08693\epsilon$ (cyan line corresponds to the cyan line in Fig. 1b for $\epsilon_Y=0.08\epsilon$), $\epsilon_Y=0.1064\epsilon$ (yellow line corresponds to yellow line in Fig. 1b for $\epsilon_Y=0.1\epsilon$). For black dotted line $\epsilon_Y=0.1637\epsilon$ and for black dashed line $\epsilon_Y=0.2321\epsilon$. The percolation lines are shown by green, red, and blue broken lines, the color code as before.}
	\label{Fig1_hvozd.pdf}
\end{figure*}


Numerical results are presented for 7-bead model of monoclonal antibody molecule to form either a neat one-component fluid ($\eta_0$=0) or to be embedded in a random array of the hard-sphere obstacles with packing fraction $\eta_0$, defined as $\eta_0=\pi\rho_0\sigma_0^3/6$. 
{The energetic parameters are for the AA=BB=AB pairs equal to $\epsilon_{}$, for the CC pairs they are zero, and for the AC=BC interaction equal to $\epsilon_{}$, 0.8$\epsilon_{}$, and 0.7$\epsilon_{}$, as explained in caption to the Figure 1. The strength of the Yukawa attraction between each mAbs bead center and the obstacle center is denoted by $\epsilon_Y$. Another parameter to be defined is $\tau=\sigma_{1}/\sigma_0$, that is ratio between the diameter of the mAbs bead and diameter of the obstacle. All the results are given in reduced units, where $\rho^*_1=7\rho_1\sigma_{1}^3$ and $T^*=k_{B}T/\epsilon$ with $k_B$ as Boltzmann’s constant.}


In the previous calculation~\cite{Hvozd2020}, we investigated effects of the antibody--obstacle repulsion on the colloidal stability of protein solution. In the present contribution the model is extended; besides the hard-sphere repulsion the obstacles can attract antibodies with a 
Yukawa potential, as further explained in Section 2. The effect of strength of this attractive interaction is examined.

\subsection{Liquid-liquid phase separation and percolation lines}

The liquid-liquid phase separation is the most important property examined in this study. The results are presented in Fig. 1(a)--(d) for different values of the antibody-antibody (labeled on the figures) and antibody-obstacle attractive interaction (denoted by $\epsilon_Y$ values in captions to figures). 

There are two curves on Figure 1 which need to be explained before the others. One is the green one; it applies to the calculation for a neat mAbs fluid (no obstacles present), where $\eta_0$=0. The second such curve is colored blue. It denotes the results obtained for the case where hard sphere obstacles (no protein-obstacle attraction; $\epsilon_Y=0$) are present. First three figures (Fig. 1a,b,c) all apply to $\tau$=1.0, i.e. the situation where the size of the mAbs bead is equal to the size of the obstacle.  

The phase boundaries for these two cases are important to further discuss the influence of crowding to the phase transition. As wee see (blue lines) the effect of the hard-sphere obstacles is to make the liquid-liquid phase boundary narrower, this is indicated by the transition from green to blue curve. However, it is important to stress, that the global shape of the curves is similar. They both, below a certain value of the reduced temperature T*, which depends on the AC=BC interaction, assume a constant $\eta$ value. 

Consider next the plots labeled 1a (left top figure), with the antibody-antibody interaction as indicated on the figure. An addition of attraction to obstacle-antibody interaction, see the purple and red curves with $\epsilon_Y=0.015\epsilon$ and $\epsilon_Y=0.03\epsilon$, increases the wideness of the phase boundary moving the right branch in direction toward the green line again. For some values of model parameters (see black and red curves) we see that, when approaching from high to low reduced  temperature, the model fluid enters from one phase domain into the 2-phase domain. By further decrease of T*, the model system re-enters into the one phase region again. 
We can explain this effect as a result of competing interactions between the antibodies themselves and antibodies with matrix particles.  %
Interestingly, at strong enough antibody-obstacle attraction (see, for example, yellow curves on Figure 1a and b, or the black one on Figure 1c) the phase transition boundaries approach to each other. Densities of these two coexisting phases are low and close to each other. 
For Figure 1c no results for $\epsilon_Y=0.1\epsilon$ (or above this value) could be obtained due to the convergence problems in trying to obtain the numerical solution. 

{On the same figure (Figure 1) we also present the percolation lines for  $\eta_0=0.0$, $\eta_0=0.1$, $\epsilon_Y=0.0$, $\epsilon_Y=0.015\epsilon$, $\epsilon_Y=0.03\epsilon$, $\epsilon_Y=0.06$, $\epsilon_Y=0.08\epsilon$ and, $\epsilon_Y=0.1\epsilon$. The color code is the same as for the other plots, we denote the data for broken lines. As we see, the strength of the AC=BC interaction, as also the strength of the Yukawa attraction between the antibodies and matrix particles, do not make much of the influence on these curves. The calculations for Fig. 1d practically coincide with those shown on plot Fig. 1b and the results are not shown here. 
}

The situation above resembles the, so called, ``empty liquid'' case, first described by Bianchi and co-workers\cite{Bianchi2006PRL}. In their paper they varied the number of sticky points (decreasing it) to make the spinodal curves narrower (cf. Figure 4 of their paper).  In our case we let the number of sticky points fixed but we vary the strength of the attraction between proteins and obstacles. In this way we influence the dimensionality of the clusters, what creates this kind of effects. Bianchi et al also found that thermodynamic perturbation theory of Wertheim \cite{Wertheim2,Wertheim3} is a reliable source of information, leading good agreement with computer simulations.
The reduced critical temperature appears to be the highest for the neat antibody fluid (green line) and the effect of adding a repulsive only hard sphere obstacle (blue line) yields a decrease of the critical temperature and density as known from before\cite{Hvozd2020}. However, the ordering of the critical values with increasing $\epsilon_Y$ interaction is not monotonic and it depends also on the antibody-antibody interaction as also on the ratio of the size of the particles involved, as shown next.

The last result in this series of calculations is illustrated in Figure 1d. Here the data are presented showing the effect of the ratio of the bead (mAbs) and obstacle size $\tau=\sigma_{hs}/\sigma_0$ = 0.5. This way we wish to get an idea of the effect of the bead/obstacle size ratio. 
%

%

The results indicate that larger size obstacles make the phase transition region considerably wider (compare the yellow lines in 1b and 1d), but other observations and qualitative trends are the same as discussed before for plots 1a and 1b.

\section{Conclusions}

Numerical results for the model antibody fluid  distributed within an array of spherical obstacles, which may attract antibodies, yield the following conclusions: (i) for the obstacles characterized merely by the excluded volume (no antibody-obstacle attraction), the phase boundary curve (spinodal line) of the model monoclonal antibody fluid becomes more narrow (see the transition from green to blue line in Figure 1) as it is for the no obstacle case. (ii) This effect is partially offset by introducing the attractive protein-obstacle interaction as in Figure 1a indicated by the red, black, and cyan lines. (iii) The black line (as also the cyan and yellow on the 1a graph) exhibits the re-entering behaviour as do also  blue and yellow curves. In other words, by cooling the system proceeds from one phase domain into the two phase region and than, at still lower temperature, it returns to the one phase region again. (iv) This effect is a consequence of the changed ratio between the 3- and 2-dimensional (linear) clusters, caused by the obstacle-mAbs attraction. 
(iv) At the same time we observe the narrowing of the spinodal curve and approach to the ``empty liquid '' situation.  
\bigskip

\section{Appendix}
	\setcounter{equation}{0}
	\renewcommand{\theequation}{A\arabic{equation}}
	\subsection{Expressions for Helmholtz free energy, chemical potential and pressure of the
		hard-sphere fluid confined in the hard-sphere matrix}
	According to SPT~\cite{Holovko2009,Patsahan2011,Holovko2013,Kalyuzhnyi2014a,Hvozd2017,Hvozd2018,Nelson2020} we have 
	\begin{equation}
		{\beta\Delta A_{hs}\over N_{hs}}=\beta\Delta\mu_{hs}-{\beta \Delta P_{hs}\over\rho_{hs}}
		\label{Ahs}
	\end{equation}
	and
	\begin{equation}
		g_{hs}={1\over \phi_0-\eta}+{3(\eta+\eta_0\tau)\over 2(\phi_0-\eta_0)^2},
		\label{g}
	\end{equation}
	where $N_{hs}$ is the number of hard-sphere particles, i.e. $N_{hs}=7N$, $N$ is the number of the antibody molecules in the system, 
	$$
	{\beta\Delta P_{hs}\over\rho_{hs}}={1\over 1-\eta/\phi_0}{\phi_0\over\phi}
	+\left({\phi_0\over\phi}-1\right){\phi_0\over\eta}\ln\left(1-{\eta\over\phi_0}\right)
	$$
	\begin{equation}
		+{a\over 2}{\eta/\phi_0\over(1-\eta/\phi_0)^2}+
		{2b\over 3}{(\eta/\phi_0)^2\over(1-\eta/\phi_0)^3}-1,
		\label{P}
	\end{equation}
	$$
	\beta\Delta\mu_{hs}=
	\beta\mu^{(ex)}_{1}
	-\ln\left(1-{\eta\over\phi_0}\right)
	+
	{\eta(\phi_0-\phi)\over\phi_0\phi(1-\eta/\phi_0)}
	+\left(1+a\right){\eta/\phi_0\over(1-\eta/\phi_0)}
	$$
	\begin{equation}
		+{(a+2b)\over 2}{(\eta/\phi_0)^2\over(1-\eta/\phi_0)^2}
		+{2b\over 3}{(\eta/\phi_0)^3\over(1-\eta/\phi_0)^3},
		\label{mu}
	\end{equation}
	and $\eta_0=\pi\rho_0\sigma_0^3/6$, $\phi_0=1-\eta_0$, $\eta=\pi\rho_{hs}\sigma_{hs}^3/6$
	and $\phi=\exp{(-\beta\mu^{(ex)}_{1})}$.
	
	\noindent
	Here
	\begin{equation}
		a=6+{3\eta_0\tau\left(\tau+4\right)\over 1-\eta_0}+
		{9\eta_0^2\tau^2\over(1-\eta_0)^2},
		\;\;\;\;\;\;\;\;\;\;\;\;
		b={9\over 2}\left(1+{\tau\eta_0\over 1-\eta_0}\right)^2,
		\label{b1}
	\end{equation}
	$$
	\beta\mu_{1}^{(ex)}=-\ln{(1-\eta_0)}+{9\eta_0^2\over 2(1-\eta_0)^2}-\eta_0Z_0
	+\left[3\eta_0Z_0-{3\eta_0(2+\eta_0)\over(1-\eta_0)^2}\right](1+\tau)
	$$
	\begin{equation}
		-\left[3\eta_0Z_0-{3\eta_0(2+\eta_0)\over 2(1-\eta_0)^2}\right](1+\tau)^2
		+\eta_0Z_0(1+\tau)^3,
		\label{mu1}
	\end{equation}
	$Z_0=(1+\eta_0+\eta_0^2)/(1-\eta_0)^3$ and $\tau=\sigma_{hs}/\sigma_0$.\\
	
	\subsection{Replica Ornstein-Zernike equations and their closures}
	
	The set of the ROZ equations enables one to calculate the structure and thermodynamic properties of the fluid adsorbed into the disorder porous media \cite{Given1992,Given1993}. 
	For  hard-sphere fluid confined in the Yukawa hard-sphere matrix the theory is represented 
	by the OZ equation for the direct $c_{00}(r)$ and total $h_{00}(r)$ correlation functions,
	describing the structure of the matrix, i.e.
	\be
	h_{00}-c_{00}=\rho_0c_{00}\otimes h_{00},
	\label{MM}
	\ee
	and a set of three equations, which include direct $c_{01}(r),c_{11}(r),c_{11(1)}(r)$ and total
	$h_{01}(r),h_{11}(r),h_{11(1)}(r)$ matrix-fluid (with the lower indices 01) and fluid-fluid 
	(with the lower indices 11 and 11(1)) correlation functions, 
	
	\be
	h_{01}-c_{01}=\rho_0c_{01}\otimes h_{00}+\rho_1c_{11(1)}\otimes h_{10},
	\label{01}
	\ee
	$$
	h_{11}-c_{11}=\rho_0c_{10}\otimes h_{01}+\rho_1c_{11(1)}\otimes h_{11}
	$$
	\be
	+\rho_1\left[c_{11}-c_{11(1)}\right]\otimes h_{11(1)},
	\label{11}
	\ee
	\be
	h_{11(1)}-c_{11(1)}=\rho_1c_{11(1)}\otimes h_{11(1)},
	\label{111}
	\ee
	where the lower index $11(1)$ denote connectedness correlation functions,
	the symbol $\otimes$ denotes convolution.
	This set of equations have to be supplemented by the corresponding closure relations. We are
	using here hypernetted chain (HNC) approximation for the OZ equation (\ref{MM}) and Mean Spherical
	Approximation (MSA) for the set of equations (\ref{01}), (\ref{11}) and (\ref{111}), i.e. 
	\be
	h_{00}(r)+1=e^{-\beta U_{hs}(r)+h_{00}(r)-c_{00}(r)}
	\label{PY}
	\ee
	and
	\be
	c_{11}(r)=\left[t_{11}(r)+1\right]f_{11}^{(hs)}(r)
	\label{MSA1}
	\ee
	\be
	c_{01}(r)=-\beta U^{(Y)}_{01}(r)e_{01}^{(hs)}(r)+\left[t_{01}(r)+1\right]f_{01}^{(hs)}(r)
	\label{MSA0}
	\ee
	\be
	c_{11(1)}(r)=\left[t_{11}(r)+1\right]f_{11}^{(hs)}(r)
	\label{MSA2}
	\ee
	
	Solution of the set of equations (\ref{MM}) - (\ref{111}) was obtained
	numerically via direct iteration method.  Yukawa contribution to Helmholtz free energy 
	$\Delta A_Y$ was calculated using the energy route, i.e.
	\be
	\beta\Delta A_Y= \int_0^\beta E_Yd\beta',
	\label{AY1}
	\ee
	where for the excess internal energy $E_Y$ we have
	\be
	{\beta E_Y\over V}=2\pi\beta\rho_1\rho_0\int_0^\infty r^2
	U^{(Y)}_{01}(r)g_{01}(r)dr,
	\label{EY}
	\ee
	
	Corresponding contributions to the chemical potential and pressure are calculated using the 
	standard thermodynamic relations.

\section*{Conflicts of interest}
There are no conflicts to declare.

\section*{Acknowledgements}
This study was supported by the Slovenian Research Agency fund (ARRS) through the Program 0103--0201 and the projects J1--1708 and J1-2471. {TH and YVK acknowledge partial support of this work by the National Academy of Sciences of Ukraine, Project K$\Pi$KBK 6541230. 
MH gratefully acknowledges financial support from
the National Research Foundation of Ukraine (project No.
2020.02/0317)}





\bibliography{sample-2} 
\bibliographystyle{rsc} 


\end{document}